\documentclass[prd,a4,superscriptaddress,nofootinbib]{revtex4}
\usepackage{hyperref,amsmath,amssymb,epsfig,bm}
\usepackage{ifthen}

\begin{document}

\renewcommand{\eprint}[1]{\href{http://arxiv.org/abs/#1}{#1}}
\renewcommand{\bibinfo}[2]{\ifthenelse{\equal{#1}{isbn}}{%
\href{http://cosmologist.info/ISBN/#2}{#2}}{#2}}
\newcommand{\adsurl}[1]{\href{#1}{ADS}}

\newcommand{\Msun}{M_\odot}
\newcommand{\vrms}{v_{\text{rms}}}
\newcommand{\tot}{{\text{tot}}}
\newcommand{\ud}{{\text{d}}}
\newcommand{\Mpc}{\text{Mpc}}
\newcommand{\half}{{\textstyle \frac{1}{2}}}
\newcommand{\third}{{\textstyle \frac{1}{3}}}
\newcommand{\numfrac}[2]{{\textstyle \frac{#1}{#2}}}
\newcommand{\ra}{\rangle}
\newcommand{\la}{\langle}
\renewcommand{\d}{\text{d}}
\newcommand{\grad}{\nabla}
\newcommand{\km}{\rm{\,km\,}}
\newcommand{\rv}{r_{200}}

\newcommand{\begm}{\begin{pmatrix}}
\newcommand{\enm}{\end{pmatrix}}

\newcommand{\threej}[6]{{\begm #1 & #2 & #3 \\ #4 & #5 & #6 \enm}}
\newcommand{\fsky}{f_{\text{sky}}}
\newcommand{\arcmin}{\text{arcmin}}
\newcommand\Tr{{\rm Tr}}
\newcommand{\cla}{\mathcal{A}}
\newcommand{\clb}{\mathcal{B}}
\newcommand{\clc}{\mathcal{C}}
\newcommand{\cle}{\mathcal{E}}
\newcommand{\clf}{\mathcal{F}}
\newcommand{\clg}{\mathcal{G}}
\newcommand{\clh}{\mathcal{H}}
\newcommand{\cli}{\mathcal{I}}
\newcommand{\clj}{\mathcal{J}}
\newcommand{\clk}{\mathcal{K}}
\newcommand{\cll}{\mathcal{L}}
\newcommand{\clm}{\mathcal{M}}
\newcommand{\cln}{\mathcal{N}}
\newcommand{\clo}{\mathcal{O}}
\newcommand{\clp}{\mathcal{P}}
\newcommand{\clq}{\mathcal{Q}}
\newcommand{\clr}{\mathcal{R}}
\newcommand{\cls}{\mathcal{S}}
\newcommand{\clt}{\mathcal{T}}
\newcommand{\clu}{\mathcal{U}}
\newcommand{\clv}{\mathcal{V}}
\newcommand{\clw}{\mathcal{W}}
\newcommand{\clx}{\mathcal{X}}
\newcommand{\cly}{\mathcal{Y}}
\newcommand{\clz}{\mathcal{Z}}
\newcommand{\CMBFAST}{\textsc{cmbfast}}
\newcommand{\CAMB}{\textsc{camb}}
\newcommand{\Omtot}{\Omega_{\mathrm{tot}}}
\newcommand{\Omb}{\Omega_{\mathrm{b}}}
\newcommand{\Omc}{\Omega_{\mathrm{c}}}
\newcommand{\Omm}{\Omega_{\mathrm{m}}}
\newcommand{\omb}{\omega_{\mathrm{b}}}
\newcommand{\omc}{\omega_{\mathrm{c}}}
\newcommand{\omm}{\omega_{\mathrm{m}}}
\newcommand{\Omdm}{\Omega_{\mathrm{DM}}}
\newcommand{\Omnu}{\Omega_{\nu}}
\newcommand{\vpsi}{\mathbf{\psi}}
\renewcommand{\vr}{\mathbf{r}}

\newcommand{\vTheta}{\mathbf{\Theta}}
\newcommand{\vdelta}{\boldsymbol{\delta}}

\newcommand{\vU}{\mathbf{U}}
\newcommand{\vQ}{\mathbf{Q}}

\newcommand{\Oml}{\Omega_\Lambda}
\newcommand{\OmK}{\Omega_K}

\newcommand{\Hunit}{~\text{km}~\text{s}^{-1} \Mpc^{-1}}
\newcommand{\Gyr}{{\rm Gyr}}
\newcommand{\muK}{\mu\rm{K}}
\newcommand{\muKarcmin}{\,\muK\,\arcmin}

\newcommand{\nrun}{n_{\text{run}}}

\newcommand{\lmax}{l_{\text{max}}}

\newcommand{\zre}{z_{\text{re}}}
\newcommand{\mpl}{m_{\text{Pl}}}

\newcommand{\valpha}{{\boldsymbol{\alpha}}}
\newcommand{\vgrad}{{\boldsymbol{\nabla}}}

\newcommand{\vphi}{\mathbf{\psi}}
\newcommand{\vv}{\mathbf{v}}
\newcommand{\vd}{\mathbf{d}}
\newcommand{\vC}{\mathbf{C}}
\newcommand{\vT}{\mathbf{\Theta}}
\newcommand{\vX}{\mathbf{X}}
\newcommand{\vn}{\mathbf{n}}
\newcommand{\vy}{\mathbf{y}}
\newcommand{\mN}{\bm{N}}
\newcommand{\eV}{\,\text{eV}}
\newcommand{\vtheta}{\bm{\theta}}
\newcommand{\tT}{\tilde{T}}
\newcommand{\tE}{\tilde{E}}
\newcommand{\tB}{\tilde{B}}

\newcommand{\mCh}{\hat{\bm{C}}}
\newcommand{\Ch}{\hat{C}}

\newcommand{\Bt}{\tilde{B}}
\newcommand{\Et}{\tilde{E}}
\newcommand{\bld}[1]{\mathrm{#1}}
\newcommand{\mLambda}{\bm{\Lambda}}
\newcommand{\mA}{\bm{A}}
\newcommand{\mC}{\bm{C}}
\newcommand{\mQ}{\bm{Q}}
\newcommand{\mU}{\bm{U}}
\newcommand{\mX}{\bm{X}}
\newcommand{\mV}{\bm{V}}
\newcommand{\mP}{\bm{P}}
\newcommand{\mR}{\bm{R}}
\newcommand{\mW}{\bm{W}}
\newcommand{\mD}{\bm{D}}
\newcommand{\mI}{\bm{I}}
\newcommand{\mH}{\bm{H}}
\newcommand{\mM}{\bm{M}}
\newcommand{\mS}{\bm{S}}
\newcommand{\mzero}{\bm{0}}
\newcommand{\mL}{\bm{L}}

\newcommand{\btheta}{\bm{\theta}}
\newcommand{\bphi}{\bm{\psi}}

\newcommand{\vb}{\mathbf{b}}
\newcommand{\vA}{\mathbf{A}}
\newcommand{\vAt}{\tilde{\mathbf{A}}}
\newcommand{\ve}{\mathbf{e}}
\newcommand{\vE}{\mathbf{E}}
\newcommand{\vB}{\mathbf{B}}
\newcommand{\vEt}{\tilde{\mathbf{E}}}
\newcommand{\vBt}{\tilde{\mathbf{B}}}
\newcommand{\vEw}{\mathbf{E}_W}
\newcommand{\vBw}{\mathbf{B}_W}
\newcommand{\vx}{\mathbf{x}}
\newcommand{\vXt}{\tilde{\vX}}
\newcommand{\vXb}{\bar{\vX}}
\newcommand{\vTb}{\vTheta}
\newcommand{\vTt}{\tilde{\vT}}
\newcommand{\vY}{\mathbf{Y}}
\newcommand{\vBwr}{{\vBw^{(R)}}}
\newcommand{\RW}{{W^{(R)}}}

\newcommand{\mUt}{\tilde{\mU}}
\newcommand{\mVt}{\tilde{\mV}}
\newcommand{\mDt}{\tilde{\mD}}

\newcommand{\Rot}{\begm \mzero &\mI \\ -\mI & \mzero \enm}
\newcommand{\Pt}{\begm \vEt \\ \vBt \enm}

\newcommand{\edth}{\,\eth\,}
\renewcommand{\beth}{\,\overline{\eth}\,}

\newcommand{\sE}{{}_{|s|}E}
\newcommand{\sB}{{}_{|s|}B}
\newcommand{\sElm}{\sE_{lm}}
\newcommand{\sBlm}{\sB_{lm}}
\newcommand{\vnhat}{{\hat{\mathbf{n}}}}

\newcommand{\vk}{{\mathbf{k}}}
\newcommand{\vl}{{\mathbf{l}}}
\newcommand{\vlhat}{{\hat{\mathbf{l}}}}
\newcommand{\vL}{{\mathbf{L}}}
\newcommand{\vq}{{\mathbf{q}}}

\title{Cluster Masses from CMB and Galaxy Weak Lensing}

\author{Antony Lewis}
\homepage{http://cosmologist.info}

\author{Lindsay King}
 \affiliation{Institute of Astronomy, Madingley Road, Cambridge, CB3 0HA, UK.}

\begin{abstract}
Gravitational lensing can be used to directly constrain the projected density profile of galaxy clusters. We discuss possible future constraints using lensing of the CMB temperature and polarization, and compare to results from using galaxy weak lensing. We model the moving lens and kinetic SZ signals that confuse the temperature CMB lensing when cluster velocities and angular momenta are unknown, and show how they degrade parameter constraints. The CMB polarization cluster lensing signal is $\sim 1\muK$ for massive clusters and challenging to detect; however it should be significantly cleaner than the temperature signal and may provide the most robust constraints at low noise levels. Galaxy lensing is likely to be much better for constraining cluster masses at low redshift, but for clusters at redshift $z\agt 1$ future CMB lensing observations may be able to do better.
\end{abstract}

\maketitle

\section{Introduction}

The distribution of clusters of galaxies as a function of mass and redshift depends on the cosmological model, and can be modelled increasingly accurately~\cite{Birkinshaw:1998qp,Shaw:2005dy,Battye:2003bm,Majumdar:2003mw}. Observations of clusters can therefore be used to learn about cosmology, as well as to test models for cluster formation and evolution.
Observations of the thermal Sunyaev-Zel'dovich (SZ) effect~\cite{SZ,Birkinshaw:1998qp,Battye:2003bm} are a powerful probe of the cluster gas, but do not measure the mass directly. To relate the gas properties to the total mass involves modelling potentially complicated baryonic gas physics. By contrast gravitational lensing probes the projected total mass, not just the gas, and therefore can provide direct information about cluster masses. In this paper we analyse the potential for reconstruction of parameterized cluster profiles from future observations of cluster lensing of the CMB and weak lensing of distant galaxies. For the first time we include constraints from CMB polarization, and also include a model of the moving lens effect that confuses the CMB temperature signal when the clusters have unknown velocities. The CMB polarization signal is much cleaner than the temperature at low noise levels, and may prove to be a good way to constrain cluster masses at high redshift. We perform an essentially optimal statistical analysis in the approximation that the unlensed fields can be treated as Gaussian.

Current CMB observations are not of high enough resolution or sensitivity to measure the cluster lensing signal. However, future missions aimed at detecting small levels of primordial gravitational waves via their distinct $B$-mode polarization signal will require both high sensitivity and resolution. This is because lensing by large scale structure can convert scalar $E$-modes into $B$-modes, and hence this lensing signal has to be subtracted to extract a small primordial $B$-mode signal from gravitational waves. The lensing reconstruction requires high resolution observations in order to have enough information to solve for both the primordial $B$-modes and the unknown large scale structure distribution~\cite{Seljak:2003pn}. It is therefore of interest to see what other useful information can be gained from such future high resolution observations. The resolution required for $B$-mode cleaning is probably rather less than needed for good cluster mass constraints from CMB lensing, however it is clearly of interest to see what can be gained from observations with slightly higher arcminute-level resolution. Cluster lensing of CMB temperature and polarization that we consider here is one potentially useful possibility.

Cluster lensing also generates a shear field that is observable by looking at the shapes of galaxies lying behind the cluster. This method of cluster mass constraint is promising and  possible with current observations, though at some level has to be limited by the finite number of source galaxies available behind the lens. The seminal paper of  Kaiser \& Squires~\cite{Kaiser:93} describes how to do a non-parametric mass reconstruction; this technique and its variants have been applied to numerous clusters (e.g. Refs.~\cite{Fischer:1997si,Clowe:2000iq}). Here we focus on parameterized cluster models, which enable statistical comparisons to be made between clusters and as a function of observational strategy.
The likelihood techniques are based on those developed in Refs.~\cite{Schneider:1999ch,King:2000wc}. We shall investigate how galaxy lensing compares to CMB lensing as a function of noise level, cluster redshift and galaxy number count. We use natural units where the speed of light is unity.

\section{CMB lensing}

\subsection{CMB temperature lensing}

\begin{figure}
\begin{center}
\psfig{figure=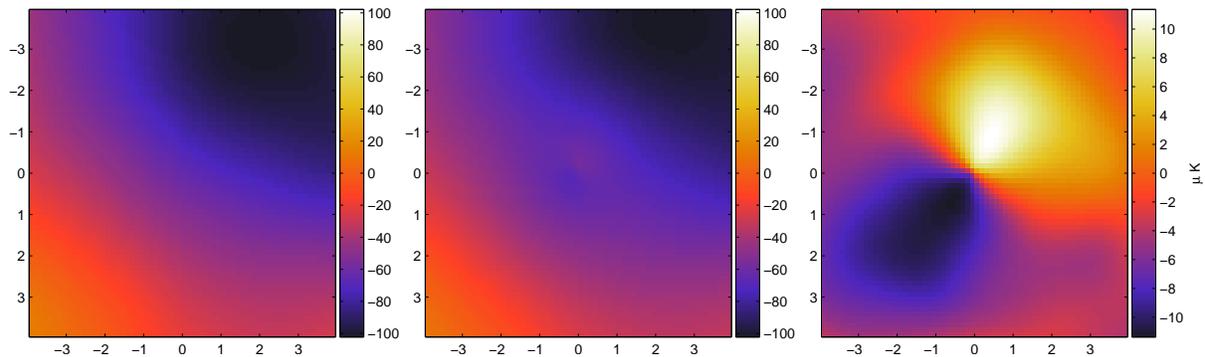,width=16cm}
\caption{Simulated effect of cluster lensing on the CMB temperature. Left: the unlensed CMB; middle: the lensed CMB; right: the difference due to the cluster lensing. The cluster is at $z=1$, and has a spherically symmetric NFW profile with mass of $M_{200} = 10^{15} h^{-1} \Msun$ and concentration parameter $c=5$. Distances are in arcminutes, and can be compared to $\rv = 3.3\,\arcmin$. This is a rather clean realization; in general the dipole pattern can be weaker and/or more complicated. Note the inverted direction of the gradient within the arcminute-scale Einstein radius in the middle figure.
\label{clustermassT}}
\end{center}
\end{figure}

The unlensed CMB is very smooth on small scales due to diffusion damping, so the small scale unlensed CMB can be locally approximated as a gradient.  Clusters act as converging lenses,  making CMB photons appear to originate further from the centre of the cluster than they actually do. So the side of the cluster on the cold side of the gradient will look hotter after lensing, and that on the hot side will look colder, giving a distinctive dipole-like signature aligned with the direction of the background CMB gradient.
 The CMB lensing signature therefore consists of small scale wiggles in the observed temperature (and polarization) in an otherwise smooth background~\cite{Seljak:1999zn,Zaldarriaga:2000ud,Dodelson:2004as,Holder:2004rp,Vale:2004rh}.
 A particular example is shown in Fig.~\ref{clustermassT}. Note that within the Einstein radius the lensing is not strictly weak (though deflection angles remain small), however we shall include the signal everywhere as the strong lensing signal on the CMB is no more difficult to model than the weak signal in the single thin lens approximation that we use.

 For a Gaussian unlensed temperature field $\Theta$, the temperature gradient variance is given by
\begin{equation}
 \la |\grad \Theta|^2\ra = \sum_l l(l+1) \frac{2l+1}{4\pi}  C_l^\Theta,
\end{equation}
where $C_l^\Theta$ is the unlensed CMB temperature power spectrum. For typical $\Lambda$CDM models that we consider here the rms is $\sim 14 \muK /\arcmin$. Since massive clusters can give deflections of the order of an arcminute, the signal is expected to be at the $\sim 10\muK$ level. The scale of the dipole-like pattern induced by the cluster lensing is much smaller than the scale of fluctuations in the unlensed CMB, and so should in principle easily be observable with high enough resolution and low enough noise. However the signal depends on the background gradient, so only clusters in front of a significant gradient can have their mass constrained this way. The gradient at a point is a Gaussian random variable, so how often this happens will depend on how sensitive the observations are to small signals and the level of complicating signals acting as sources of correlated noise.

The observed direction of a point on the CMB last scattering surface is related to the direction it would have had without lensing by a deflection angle $\valpha$, determined in the small-angle Born approximation by
\begin{equation}
\valpha(\hat{\vn}) = -2 \int_0^{\chi_S} \ud \chi  \left( 1- \frac{\chi}{\chi_S}\right) \vgrad_\perp \Psi(\chi \hat{\vn};\eta),
\label{deflect}
\end{equation}
where $\hat{\vn}$ is the direction of observation, $\chi_S$ is the comoving distance to the source at the last scattering surface (taken to be thin), $\eta$ is the time at which the photon was at position $\chi\hat{\vn}$, and $\Psi$ is the Newtonian potential. For cluster lensing the integral is dominated by the small part through the thin cluster and the angular factor $ 1- \chi_L/\chi_S$ may be taken out of the integral, where $\chi_L$ is the distance to the cluster. The potential is related to the comoving density perturbation via the Poisson equation. For a more detailed review of CMB lensing see Ref.~\cite{Lewis:2006fu}.

If the background gradient could be measured cleanly away from the cluster, the cluster deflection angles could be reconstructed directly, and hence used to solve for the cluster profile and mass given certain assumptions. Unfortunately the situation is more complicated because the unlensed CMB isn't \emph{exactly} a gradient --- clusters have finite angular size so the unlensed CMB in reality has more complicated spatial structure. There is also additional small scale power due to other lensing sources along the line of sight, not to mention other important non-linear effects.  Fortunately the problem is statistically straightforward if we take the unlensed CMB field to be Gaussian. For a given deflection field the lensed CMB is also Gaussian since the deflections just re-map points:
a uniform sampling of the lensed CMB just corresponds a non-uniform (and possibly multiple) sampling of the unlensed CMB. The correlation function of the observed temperature is therefore just given by the correlation function at the undeflected position on the last scattering surface (neglecting complications due to other lensing along the line of sight). We can therefore work out the likelihood of any given cluster deflection field $\valpha(\theta)$ for some set of cluster parameters $\theta$ using
\begin{equation}
-2 \log P(\valpha(\theta)| \tilde{\vTheta}) = \tilde{\Theta}(\vx_i) C^{-1}(\vx_i,\vx_j) \tilde{\Theta}(\vx_j) + \log |C(\vx_i,\vx_j)|,
\label{Tlike}
\end{equation}
where $i,j$ index the different observed positions (taken here to be pixel centres) which are summed over implicitly in the first term.
Here $\mC$ is given by the pixel noise covariance plus the covariance matrix determined by the correlation function of the unlensed temperature field $\Theta$:
\begin{equation}
C_\Theta(\vx,\vx') \equiv \la \Theta(\vx)\Theta(\vx')\ra = C_\Theta(\beta) = \sum_l \frac{2l+1}{4\pi} C_l^\Theta P_l(\cos\beta),
\end{equation}
where $\beta$ is the angular separation between $\vx$ and $\vx'$.
Here we only consider isotropic white (uncorrelated) noise $\sigma_N^2(\vx)$ so that
\begin{equation}
C(\vx_i,\vx_j) = C_\Theta(\vx_i+\valpha_i,\vx_j+\valpha_j) + \delta_{ij}\sigma_N^2(\vx_i).
\end{equation}
Non-zero noise regularizes the inverse, for example the case when two observations on the Einstein ring are actually sampling the same point on the last scattering surface.
The effect of lensing by other perturbations along the line of sight can be crudely modelled by using the lensed power spectrum, though a more careful analysis would require a study of the non-Gaussian distribution numerically using simulations. We shall only consider small pixelized observations, in which case the dimension of the observed data vector is quite manageable, and the covariance matrix can be inverted exactly for each set of cluster model parameters considered. For discussion of beam issues see Section~\ref{beam}.

Unfortunately the lensing cannot be observed directly, only its mixture with numerous other non-linear signals including the Sunyaev-Zel'dovich (SZ)~\cite{SZ,Birkinshaw:1998qp,Battye:2003bm} and Rees-Sciama~\cite{ReesSciama} effects.
Thermal SZ is in principle not a problem because of its distinctive spectral signature, and indeed could be used to help model the kinetic signal due to their correlated sources. Kinetic SZ is probably the most problematic~\cite{Vale:2004rh}, and has the same spectrum as the lensing signal we are interested in. For a cluster that has circular symmetry about the line of sight, the kinetic SZ signal should also have circular symmetry, and therefore be orthogonal to the dipole-like lensing signal.  However there will in general be non-symmetric spatially varying kinetic SZ both from cluster internal motion and other gas along the line of sight that is more problematic. Cluster motion transverse to the line of sight can also give a kinetic Rees-Sciama signal, and secondary signals from outside the cluster can contribute additional sources of noise.

We can easily include a known non-lensing secondary contribution $\vTheta_m$ into the posterior distribution,
\begin{equation}
-2 \log P(\theta| \tilde{\vTheta}^\tot,\vTheta_m) =(\tilde{\vTheta}^\tot -\vTheta_m)^\dag \mC^{-1} (\tilde{\vTheta}^\tot-\vTheta_m) + \log |\mC|
\label{Tlike2},
\end{equation}
where $\tilde{\vTheta}^\tot$ is the observed temperature (a sum of the lensing signal and other secondary signals). In general the other secondaries will also depend on the parameters $\theta$ and are hard to model.

\subsubsection*{Moving lens effect}

If the cluster has a significant velocity transverse to the line of sight there will be a moving lens signal~\cite{Birkinshaw83,Tuluie:1995ut,Aghanim:1998ux,Lewis:2006fu}: a photon passing the front side of the lens (with respect to the direction of motion) will see a weaker potential on its way into the lens than on the way out, and hence receives a net redshift. Similarly a photon on the other side will receive a net blue shift. The moving lens therefore induces a dipole-like temperature anisotropy, which is at the $\muK$ level for massive clusters~\cite{Aghanim:1998ux}. This is essentially the kinetic component of the Rees-Sciama effect (the nonlinear growth component gives a small non-dipole-like effect that we neglect). The moving lens signal can also be viewed as dipole lensing: in the rest frame of the cluster, the cluster sees a CMB dipole; lensing deflects photons so those appearing to come from the front hot side actually come from closer to the cold side, and hence appear colder, and conversely on the other side.

The moving lens dipole-like signal is easy to model~\cite{Birkinshaw83,Gurvits86}.  The temperature anisotropy induced by a small lens moving perpendicular to the line of sight with velocity $\vv_\perp$ (natural units) is at lowest order
\begin{equation}
\frac{\Delta \Theta}{\Theta} = -2 \int \ud\chi\, \vv_\perp \cdot \vgrad_\perp \Psi(\chi \hat{\vn};\eta) = \vv_\perp \cdot \vdelta\beta,
\end{equation}
where $\vdelta\beta$ is the deflection angle of the photon at the lens (related to the observed deflection angle by $\valpha = (1-\chi_L/\chi^*)\vdelta\beta$).
The moving lens signal therefore looks very similar to the dipole-like lensing signal, though unlike the lensing signal the direction of the dipole pattern is determined by the velocity rather than the (uncorrelated) background CMB gradient. The moving lens signal is a source of confusion for the static lensing signal if there is no way to measure the transverse cluster velocity independently (e.g. using CMB polarization).

If we model the cluster peculiar velocities as a 3-D Gaussian random field with rms velocity $\vrms$, the transverse velocity is also Gaussian with $\la \vv_\perp^2\ra = \la v_x^2 + v_y^2\ra = 2\vrms^2/3$. Writing $\vTheta_m = \Delta\vTheta = \Theta \vdelta\beta\cdot \vv_\perp \equiv v_x \vTheta_{m,x} + v_y \vTheta_{m,y}$ we then have
\begin{equation}
\mC_m \equiv \la \tilde{\vTheta}^\tot \tilde{\vTheta}^\tot{}^\dag \ra = \la \tilde{\vTheta} \tilde{\vTheta}^\dag + \vTheta_m \vTheta_m^\dag\ra = \mC + \frac{\vrms^2}{3}\left( \vTheta_{m,x}\vTheta_{m,x}^\dag + \vTheta_{m,y}\vTheta_{m,y}^\dag \right).
\end{equation}
Here we have assumed any correlation between the CMB and velocity is negligible so the fields are uncorrelated. Marginalized over the transverse cluster velocity the likelihood is therefore
\begin{equation}
-2 \log P(\theta| \tilde{\vTheta}^\tot) =\tilde{\vTheta}^\tot{}^\dag  \mC^{-1}_m \tilde{\vTheta}^\tot + \log |\mC_m|.
\end{equation}
Note that in general $\vrms$ is a function of the cluster parameters and redshift. The velocities of nearby clusters will also be correlated to some extent, in which case treating the velocity uncertainty as independent Gaussians would not be quite correct. We may also be interested to extract information about the transverse velocity, in which case the velocity posterior distribution could be calculated rather than marginalizing over it as we do here.

\subsubsection*{Kinetic SZ from cluster rotation}

In the ideal symmetric case adding a symmetric kinetic SZ template has no effect on parameter constraints because the affected modes are orthogonal to those sensitive to the dipole-like lensing signal. This is true even though the kinetic SZ is expected to have significantly larger amplitude. The extent to which asymmetric kinetic SZ confuses the signal really needs to be tested from simulations~\cite{Holder:2004rp,Vale:2004rh}, which indicate that in practice it is a major source of confusion.\footnote{Ref.~\cite{Maturi:2004zj} find a small effect from kinetic SZ, but they only consider the symmetric component.} Here we consider only a simple analytic model.

A simple situation that can give a dipole-like kinetic SZ signature is when the cluster is rotating~\cite{Chluba:2002es,Cooray:2001vy}. We can model this easily even in the idealized  case following Ref.~\cite{Cooray:2001vy}. We assume the cluster is rigidly rotating inside the virial radius, and that that the gas is in hydrostatic equilibrium in a non-rotating dark matter potential~\cite{Makino:1997dv}. The kinetic SZ temperature anisotropy is given by
\begin{equation}
\frac{\Delta \Theta}{\Theta} = \sigma_T\int \ud \chi \, n_e \hat{\vn}\cdot \vv,
\end{equation}
where the integral is along the line of sight in direction $\hat{\vn}$, and the gas velocity is determined from the radius, mass and the cluster angular momentum, the latter being parameterized by the dimensionless parameter $\lambda$ following Refs.~\cite{Cooray:2001vy,Bullock:2000ry}. The electron density $n_e$ is assumed to be associated with the fully ionized gas. We shall make the crude approximation that the angular momenta have a 3-D Gaussian distribution, with  $\lambda_{\text{rms}} = 0.04$ as in~\cite{Cooray:2001vy}. This rms value is broadly consistent with that found in simulations~\cite{Bullock:2000ry}. Assuming we have no knowledge of the cluster rotation, we can then marginalize over the kinetic SZ signal in exactly the same way that we did for the moving lens signal. The kinetic SZ signal from rotation gives a dipole-like pattern peaking at a few $\muK$ near the centre, but falls off rapidly at large radii (well inside the virial radius). As for the moving lens signal the direction of the dipole is expected to be uncorrelated with the background CMB gradient, and we assume it is also uncorrelated with the transverse velocity.

The components of the angular momentum transverse to the line of sight could be measured by other means, for example from the redshifts of the galaxies, so in principle it may be possible to model this signal out. However it is useful to consider it as an indicator of the level of problems caused by more general kinetic SZ signals from substructure and internal motion.

Since we wish to extract constraints from CMB lensing here rather than by modelling the (in reality complicated) kinetic SZ, we shall take the fixed fiducial parameters when calculating the kinetic SZ contribution to the covariance. If the kinetic SZ could be modelled reliably as a function of cluster parameters then parameter constraints could be improved.

\subsubsection*{Background secondaries}

On scales $l\agt 4000$ the spectrum from inhomogeneous reionization and secondary doppler signals is expected to be very roughly scale invariant with $l(l+1)C_l/2\pi \sim 5\muK^2$~\cite{Zahn:2005fn} and dominates the (cluster-free) lensed CMB power. The extra power provided by this signal actually increases the rms temperature gradient behind the cluster significantly, which tends to increase the lensing signal. This is compensated by the increased smaller scale power, acting as a source of correlated noise, and degrading parameter constraints. Neglecting non-Gaussianity and the different redshift of the source, it can be modelled simply by adding the approximately scale invariant power spectrum to the unlensed power spectrum used for computing the unlensed covariance.

In principle, discrete secondary sources behind the cluster will be sheared by the lens, so shear measurements can be used to obtain additional information about the cluster~\cite{Zaldarriaga:1998te}. This would require substantially higher resolution observations, and a detailed investigation is beyond the scope of this paper.

\subsection{CMB polarization lensing}

The CMB temperature cluster lensing signal is complicated by many sources of confusion, including sizeable kinetic SZ and moving lens signals. The lensing signal is also small if the temperature gradient behind a cluster happens to be small. Furthermore any attempted density profile reconstruction will have degeneracies because the single temperature gradient will not show up orthogonal deflection angles.
In principle CMB polarization observations can help with all of these problems.

The statistics of the polarization gradients is discussed in the appendix.  The rms gradient of each Stokes' parameter is $\sim 1\muK/\arcmin$, so a signal $\sim 1\muK$ is expected for massive clusters. On cluster scales the correlation with the temperature gradient is fairly small, at the $10\%$ level, so the directions of the gradients are close to independent. The addition of polarization data is therefore very much like adding two new temperature fields but with lower signal and slightly different CMB `noise' properties.

The polarization signal is about a factor of ten lower than the temperature signal, so detection will be challenging. However the polarization signal should be significantly cleaner as the other secondary polarization signals are generally small~\cite{Gibilisco:1997wr,Challinor:1999yz,Hu:1999vq,Valageas:2000ke,Shimon:2006hn}. The dominant frequency-independent signal is expected to be from scattering of the primordial CMB temperature quadrupole from ionized gas in the cluster. This depends linearly on the cluster optical depth ($\sim 1\%$), and typically gives a signal $\alt 0.1\muK$~\cite{Kamionkowski:1997na,Cooray:2002cb,Amblard:2004yp,Liu:2004pu}. This signal is strongly correlated across the sky for clusters at similar redshifts due to the small variation of the quadrupole within a Hubble volume, and is also fairly constant across each cluster (in contrast to the lensing dipole-like signal). It may therefore be possible to subtract it out. For spherically symmetric clusters the signal should be orthogonal to the lensing signal and hence irrelevant, so we shall neglect it here. There will however be hard to model spatial variations due to varying quadrupole and ionized gas density within the cluster that prevent this being done perfectly, but we can expect residuals to be $\ll 0.1\muK$.

The dominant frequency-dependent signal is likely to be re-scattering of anisotropic thermal SZ,  giving a frequency-dependent cluster polarization signal $\alt 0.7\muK$~\cite{Lavaux:2003qf,Shimon:2006hn} at the peak of the SZ spectrum. Multi-frequency observations should be able to subtract this out, and lower frequency observations would in any case see significantly less signal.
Polarization from scattering of the kinetic quadrupole is expected to be around ten times lower than the signal from scattering of the primordial quadrupole~\cite{SZ,Challinor:1999yz,Cooray:2002cb}, has a different frequency dependence, and should anyway be orthogonal to the lensing signal at lowest order. Contributions from cluster rotation are expected to be very small, $\sim 10^{-4} \muK$~\cite{Chluba:2002es}.
Other signals such as Faraday rotation also have a distinct spectral signature so in principle they can be separated out using multi-frequency observations. The small scale signal from inhomogeneous reionization is expected to be small~\cite{Weller:1999ch,Santos:2003jb}, and below the signal expected from other lenses. We shall simply assume that all non-lensing signals can be removed or are negligible, and roughly model the small effect of other lenses along the line of sight by using the lensed CMB polarization power spectra.

The unlensed polarization fields are expected to be Gaussian, and the full polarized posterior can be computed as for the temperature using the correlation functions between the Stokes' parameters at the undeflected positions.
The polarization field can be described as a complex spin two field $P=Q + i U$. The scalar correlation function between polarization at $\vx$ and $\vx'$ should be independent of the basis used to define $P$ at the two points. To do this, we want to describe the polarization in the physically relevant basis defined by $\vr = \vx - \vx'$. If $\vr$ makes an angle $\phi_r$ to the $\ve_x$ axis, this amounts to rotating the basis by an angle $\phi_r$ anticlockwise at each point, giving $P_r(\vx) = e^{-2i\phi_r} P(\vx)$. In this physical basis we can then define the basis-independent correlation functions~\cite{Chon:2003gx,Challinor:2005jy}
\begin{eqnarray}
 \xi_+(\beta) &\equiv& \la P_r(\vx)^* P_r(\vx') \ra = \la P(\vx)^* P(\vx') \ra =  \sum_{l'} \frac{2l+1}{4\pi}
(C_l^E + C_l^B) d^l_{22}(\beta)\\
 \xi_-(\beta) &\equiv&  \la P_r(\vx)\, P_r(\vx') \ra =  \la e^{-4i\phi_r} P(\vx) P(\vx') \ra =\sum_{l} \frac{2l+1}{4\pi}
(C_l^E - C_l^B)d^l_{2\, -2}(\beta) \\
 \xi_X(\beta)  &\equiv&  \la \Theta(\vx)\, P_r(\vx') \ra =  \la  \Theta(\vx) e^{-2i\phi_r} P(\vx') \ra = \sum_l \frac{2l+1}{4\pi}
C_l^X d^l_{02}(\beta).
\end{eqnarray}
Here $C_l^E$ and $C_l^B$ are the $E$-mode (gradient-like) and $B$-mode (curl-like) power spectra~\cite{Zaldarriaga:1996xe}, and $C_l^X$ is the cross correlation of the $E$-modes with the temperature. The $d^l_{mn}$ are the reduced Wigner functions (see e.g. Ref.~\cite{AngularMom}).
Note that a pure polarization gradient is neither unambiguously $E$ or $B$, as the decomposition is non-local and depends on second derivatives. The decomposition into $E$- and $B$-modes is not especially helpful for analysing the cluster polarization signal: on cluster scales both the unlensed $E$- and $B$-mode power spectra are expected to be small, with lensing introducing approximately equal small scale power into both (see the appendix). For Gaussian fields an optimal analysis can be performed using the Stokes parameters directly without decomposition into $E$- and $B$-modes.

Writing the observed pixel temperature, $Q$ and $U$ as a combined vector $(\vT, \vQ , \vU)$, these correlation functions are sufficient to calculate the full covariance matrix that we need for computing the likelihood function given by the obvious generalization of Eq.~\eqref{Tlike}. Since the temperature signal is likely to be complicated by secondary signals, we shall mostly consider the polarization measurements separately from the temperature, though including the full correlation structure is no problem if required. Since the polarization signal is so much smaller than the temperature, any noise level that allows polarization detection will generally give a much better constraint from the temperature \emph{if} the temperature signal is clean enough. However in practice confusion with other secondaries may make the polarization more useful at low noise levels.

\subsection{Beam and pixelization}
\label{beam}

The effect of beam convolution is complicated because the beam convolves the lensed rather than unlensed sky. An effectively circular Gaussian beam will in general measure a non-circular non-Gaussian average of the unlensed sky due to the non-uniform lensing field. If the unlensed sky and deflection angle over a pixel can be well approximated by a gradient, a pixel (or pixel-size beam) average will be very close to the value at the pixel centre, which is what we use here. However inside the Einstein radius the deflection angle gradient can change significantly on the pixel scale, and a more careful analysis would be required for very accurate results.

If the centre of a circular pixel is aligned with the centre of a spherically symmetric cluster, an integral over the pixel should give the same as the value at the centre of the pixel, which should be the same as the unlensed value. In this case the pixel gives essentially no information about the cluster. However if the grid of pixels is offset from the centre of the cluster, two adjacent central pixels will have different signals and slightly more information can be gained about the central cluster profile. This manifests itself as a somewhat better upper limit on the concentration parameter $c$ when pixels centres are offset. Since a generic observation will not be aligned exactly with the cluster we have chosen to offset our pixels so as not to introduce artificial symmetries.  However there is some error from using the values at the centre of the pixels. We find that for the basic case the results are fairly insensitive to the pixel integration (checked explicitly by calculating the covariance for pixel values taken to be an average of 4 or 9 sub-pixels). The behaviour is more complicated when there is significant small scale unlensed power, eg. from inhomogeneous reionization. We do not attempt to model the beam effect in detail here. When small scale unlensed power is included our results should be taken as an approximate estimate; a more detailed analysis would be needed for any given actual experimental beam. For distant clusters with masses $\sim 2\times 10^{14} \Msun$ the Einstein radius is around $0.2\,\arcmin$, which for aligned pixels would lie entirely within a central pixel of side $0.5\,\arcmin$. In this case the result from using aligned pixels (where the central pixel contributes no information) should give an idea of the result when the strong lensing region is ignored, which corresponds to somewhat increasing the upper limit on $c$ compared to the results we present using all the (offset) pixels.

The effect of pixel (beam) size and cluster redshift is explored in detail in Ref.~\cite{Dodelson:2004as}. Here we shall fix the pixel size to $0.5\,\arcmin$, and use a square box of side $8\,\arcmin$ (16 pixels) nearly centred over the cluster. Due to the small scale power in the CMB, using larger box sizes gains very little since the cluster signature cannot be distinguished from variations in the primordial CMB, and our results are stable to increasing the box size. We use the flat sky approximation, which should be very accurate on cluster scales. The sub-arcminute resolution we are considering is somewhat beyond the capabilities of currently planned CMB experiments, but could be achievable in future.

\section{Galaxy lensing}

Galaxies lying behind a cluster are assumed to have uncorrelated shapes. Since we do not wish to assume anything about the spatial distribution of the unlensed galaxies, observing the lensed galaxies tells us nothing directly about the deflection angle. However lensing does shear the galaxies such that the observed shapes after lensing are correlated. Specifically, any function of the shape that transforms under shear like an ellipticity will give an unbiased estimator of the shear due to lensing. Observations of the lensed galaxy shapes can therefore be used to constrain the cluster profile via the observed shear.
Lensing also modifies the number counts of galaxies behind a cluster, and this effect can also be used to probe the cluster profile. However, in practice, shear is predominantly used rather than magnification, since it requires no external calibration (see Ref.~\cite{Schneider:1999ch}).

We shall not be concerned with the details of galaxy shape measurement here, and only consider the ideal case in which the point spread function can be accounted for exactly and there are no other observational systematics. We assume some shape measurement $\epsilon =\epsilon_+ + i \epsilon_\times$ that transforms like an ellipticity, gives an unbiased estimator of the reduced shear, $\la \epsilon \ra = g$, and that it can be measured exactly. The quantity $\epsilon_+$ is an ellipticity in the direction of some chosen basis axes, and $\epsilon_\times$ is the corresponding ellipticity in a basis rotated by $45^\circ$; we work in a polar basis, centred on the cluster.
We neglect any small correlations in the unlensed galaxy shapes, and take the galaxies to be essentially point-like with known position $\vx$. The dispersion of the intrinsic shapes of the galaxies provides a source of `noise' on the shear, and means that the shear cannot be measured perfectly because there are only a finite number of lensed galaxies.

Our measure for (complex) ellipticity $\left|\epsilon\right|{\rm exp}(2i\phi)$ is such that its modulus corresponds to
$(1-r)/(1+r)$ where $r$ is the galaxy's axis ratio, and $\phi$ is its position angle.
The distribution of galaxy ellipticities is non-Gaussian; however for our purposes we shall approximate the distribution as a Gaussian with some variance $\sigma^2$, which is related to the unlensed variance $\sigma_{\rm u}^2$ through~\cite{Schneider:1999ch}
\begin{equation}
\sigma\approx\left(1-\left|g\right|^2\right)\sigma_{\rm u},
\end{equation}
which is a reasonable approximation for weak shears. The likelihood function for one galaxy is then given to within a constant by
\begin{equation}
-2 \log P(\theta| \epsilon_i,z_i,\vx_i) = 2 |\epsilon_i -g(\theta,\vx_i,z_i)|^2/\sigma^2 + 2\log \sigma^2.
\end{equation}
Here $z_i$ is the redshift of each source galaxy (we assume the cluster redshift is well known). In the case when the source redshifts are not known exactly, the marginalized probability is given by
\begin{equation}
P(\theta|\epsilon_i,\vx_i) = \int \ud z_i P(\theta|\epsilon_i,z_i,\vx_i)P(z_i)
\end{equation}
where $P(z_i)$ is the probability distribution for $z_i$. For $z_i$ less than the cluster redshift the shear is zero, so $P(\theta|\epsilon_i,z_i,\vx_i)$ is independent of $\theta$.
 Since we assume the intrinsic ellipticities are uncorrelated and neglect systematics, the probability from observations of $N$ galaxies is just the product of the probability from each.

Since these results are only valid in the weak lensing regime, we exclude the central region of the cluster around the Einstein ring; the same physical region is excised for a cluster of a particular mass. This is of the order of an arcminute for massive clusters.

\section{Expected Results}
\subsection{Mean log likelihoods}
We follow Ref.~\cite{Schneider:1999ch} by calculating the expected log likelihood for a set of cluster parameters $\theta$. For some fiducial model with parameters $\theta_0$ this is given by $\la \log P(\theta | \text{data})\ra$ where the average is over data realizations that could come from the $\theta_0$ model, given the noise properties. For observations of a large number of identical clusters, this mean log likelihood gives the average contribution of one cluster to the total joint log likelihood. For the CMB we have
\begin{equation}
-2 \la \log P(\theta|\theta_0) \ra = \Tr\left[\mC(\theta_0)\, \mC^{-1}(\theta)\right] + \log |\mC(\theta)|
\end{equation}
where $\mC(\theta)$ is the correlation matrix for the undeflected points. The mean log likelihood peaks at the true model $\theta=\theta_0$, and the shape of the likelihood contours give a good idea of the degeneracy directions expected from particular observations. For the temperature case $\mC$ is replaced with $\mC_m$ when we want to marginalize over the moving lens or rotating SZ signals.

The mean log likelihood for one galaxy at redshift $z$ is~\cite{Schneider:1999ch}
\begin{eqnarray}
-2\la \log P(\theta|\theta_0,\vx,z) \ra &=& -2\int \ud^2\epsilon P(\epsilon|\theta_0) \log P(\theta| \epsilon,z,\vx)\nonumber \\
&=&2\left( \frac{|g(\theta,\vx,z)-g(\theta_0,\vx,z)|^2 + \sigma_{0}^2}{\sigma^2} + \log \sigma^2\right)\,,
\end{eqnarray}
where $\sigma_{0}$ denotes $\sigma\left(g\left(\theta_0\right)\right)$.
Given the probability $P(N)$ to find $N$ galaxies in the data field (which follows a Poisson distribution and is independent of position) the number density observed  $n(\theta_0,\vx,z)$ depends on the magnification due to the lensing. We follow Ref.~\cite{Schneider:1999ch} by taking the scaling of number density with magnification $\mu$ to be $[\mu(\theta_0,\vx,z)]^{\beta-1}$, where $\beta$ is the slope of the unlensed number counts, taken to be 0.5.
The number density function averaged over the total number density distribution is then
\begin{equation}
\la n(\theta_0,\vx,z)\ra_N = \int \ud N [\mu(\theta_0,\vx,z)]^{\beta-1} N P(N) P(z) = n_\gamma [\mu(\theta_0,\vx,z)]^{\beta-1} P(z),
\end{equation}
where $n_\gamma$ is the expected unlensed angular number density (including all redshifts) for which a shape can be measured. The source galaxies have a redshift probability distribution $P(z)$ normalized so $\int_0^\infty \ud zP(z)=1$. The total mean log likelihood is then
\begin{eqnarray}
-2\la \log P(\theta|\theta_0) \ra &=&
-2  \int \ud z  \int \ud^2 \vx\,  \la n(\theta_0,\vx,z)\ra_N \la \log P(\theta|\theta_0,\vx,z) \ra_\epsilon\nonumber
\\
&=& 2 n_\gamma \int \ud z P(z) \int \ud^2 \vx [\mu(\theta_0,\vx,z)]^{\beta-1}\left( \frac{|g(\theta,\vx,z)-g(\theta_0,\vx,z)|^2 + \sigma_{0}^2}{\sigma^2} + \log \sigma^2\right)\,
\end{eqnarray}
as derived in Ref.~\cite{Schneider:1999ch}. Monte Carlo simulations were also performed in Ref.~\cite{Schneider:1999ch} which showed that the scatter of maximum likelihood points in different realizations corresponds well with the mean log likelihood contours.
For galaxies with redshift less than the cluster the shear is zero, and there is no contribution to the integral (except an irrelevant constant).
This result is valid when the observations measure the redshift of each galaxy accurately. In the case when the redshifts are uncertain the result is more complicated
\begin{eqnarray}
-2 \la \log P(\theta|\theta_0) \ra &=& -2 \int \ud z \int \ud^2 \vx\, \la n(\theta_0,\vx,z)\ra_N \la   \log P(\theta|\theta_0,\vx) \ra_\epsilon\nonumber \\
&=&
-2 n_\gamma \int \ud z  P(z) \int \ud^2 \vx\,  [\mu(\theta_0,\vx,z)]^{\beta-1}
\int \ud^2 \epsilon\, P(\epsilon|\theta_0,z,\vx) \log \left[\int \ud z' P(\theta|\epsilon,z',\vx)P(z'|z)\right],
\label{marge_eq}
\end{eqnarray}
which cannot easily be simplified any further. Here $P(z'|z)$ is the post-observation distribution of the redshift $z'$ given the galaxy is actually at $z$. Note that Eq.\,(20) reduces to Eq.\,(19) when the source redshifts are known.

Throughout we shall assume a purely adiabatic standard $\Lambda$CDM cosmology with Hubble parameter $H_0=70\km s^{-1} \Mpc^{-1}$, dark matter density $\Omega_c h^2=0.11$, baryon density $\Omega_b h^2=0.022$, spectral index $n_s=0.99$, primordial curvature perturbation amplitude $A_s=2.5\times 10^{-9}$ (corresponding to matter fluctuation parameter today $\sigma_8\approx 0.87$) and optical depth $\tau=0.15$.

\subsection{Spherically symmetric NFW clusters}

Our main purpose here is to compare the power of different sources for cluster mass lensing reconstruction. We therefore use a simple parameterization for the cluster radial profile, and see how well the parameters can be constrained using different methods. We shall follow Ref.~\cite{Dodelson:2004as} and assume clusters with a spherically symmetric NFW profile~\cite{NFW} given by
\begin{equation}
\rho(r) = \frac{A}{r(c r+\rv)^2}
\end{equation}
for some concentration parameter $c$, and radius $\rv$. We shall parameterize the mass of the cluster by that inside the radius $\rv$ defined so that it is 200 times the mean density of a critical density universe at the cluster redshift,
\begin{equation}
M_{200} \equiv 200 \frac{4\pi}{3} \rho_c(z) \rv^3,
\end{equation}
where $\rho_c(z)$ is the critical density at the redshift of the cluster, $3H^2(z)/8{\pi}G$. The amplitude is given in terms of the mass and concentration parameter by
\begin{equation}
A = \frac{M_{200} c^2}{4\pi[\ln(1+c) - c/(1+c)]}.
\end{equation}
The deflection at the cluster is then given by~\cite{Dodelson:2004as}
\begin{equation}
\vdelta\beta(\vr) = \frac{-16\pi G A}{c^2 \, r_s} F(r/ r_s) \hat{\vr},
\end{equation}
where $\vr$ is the transverse distance from the centre of the cluster, $\hat\vr\equiv \vr/r$,  $r_s\equiv \rv/c$ is the scale radius, and
\begin{eqnarray}
F(x<1) &=& \frac{1}{x}\left(\ln(x/2) + \frac{\ln( x/[ 1-\sqrt{1-x^2}])}{\sqrt{1-x^2} }\right) \nonumber\\
F(x=1) &=& 1-\ln(2) \\
F(x>1) &=& \frac{1}{x}\left(\ln(x/2) + \frac{\pi/2 - {\rm sin^{-1}}(1/x)}{\sqrt{x^2-1}} \right).\nonumber
\end{eqnarray}
The observed deflection is then given by $\delta\theta = (1-\chi_L/\chi_S)\delta\beta$ where $\chi_L$ and $\chi_S$ are the comoving distances to the lens and the source (for the CMB taken to be the point of maximum visibility on the last scattering surface).

The convergence $\kappa$ is (minus one half) the angular divergence of the observed deflection angle, and at the radial distance $r$ from the centre of the cluster is given by (see e.g. Ref.~\cite{Bartelmann:1996hq})
\begin{equation} \kappa(\vr)=\kappa_k f(r/r_s)\,,
\end{equation}
where $f(x) = x^{-1} \frac{\ud}{\ud x} [x F(x)]$ is
\begin{eqnarray}
f(x<1)&=&\frac{1}{x^{2}-1}\left(1-\frac{2~{{\rm tanh^{-1}}\sqrt{(1-x)/(1+x)}}}{\sqrt{1-x^{2}}}\right)\nonumber\\
f(x=1)&=&\frac{1}{3}\\
f(x>1)&=&\frac{1}{x^{2}-1}\left(1-\frac{2~{\rm tan^{-1}}\sqrt{(x-1)/(x+1)}}{\sqrt{x^{2}-1}}\right).\nonumber
\end{eqnarray}
Here
\begin{equation}
\kappa_{k}\equiv \frac{2r_{s}\rho_s}{\Sigma_{\rm crit}} = \frac{8\pi G A}{c^2 r_s^2} D_L (1-\chi_L/\chi_S),
\end{equation}
where $\Sigma_{\rm crit} = [4\pi G D_L(1-\chi_L/\chi_S)]^{-1} $ is the (redshift dependent) critical surface mass density of the lens and the angular diameter distance to the lens is $D_L$. The characteristic density of the halo $\rho_s$
is related to the amplitude $A$ through
\begin{equation}
\rho_s =\frac{Ac}{\rv^3}\,.
\end{equation}
In a polar basis centred on the cluster the shear $\gamma$ is real and given by
\begin{equation}
\gamma(\vr)=\kappa_{k}j(r/r_s)\;,
\end{equation}
where
\begin{eqnarray}
j(x<1)&=&\frac{4~{\rm
tanh^{-1}}\sqrt{(1-x)/(1+x)}}{x^{2}\sqrt{1-x^{2}}}+\frac{2\ln\left(\frac{x}{2}\right)}{x^{2}}-\frac{1}{x^{2}-1}
\frac{2~{\rm tanh^{-1}}\sqrt{(1-x)/(1+x)}}{\left(x^{2}-1\right)\sqrt{1-x^{2}}}\nonumber\\
j(x=1)&=&2\ln\left
(\frac{1}{2}\right )+\frac{5}{3}\\
j(x>1)&=&\frac{4~{\rm
tan^{-1}}\sqrt{(x-1)/(x+1)}}{x^{2}\sqrt{x^{2}-1}}+\frac{2\ln\left(\frac{x}{2}\right)}{x^{2}}-\frac{1}{x^{2}-1}
+\frac{2~{\rm tan^{-1}}\sqrt{(x-1)/(x+1)}}{(x^{2}-1)^{\frac{3}{2}}}\nonumber\,.
\end{eqnarray}
The reduced shear $g$ follows from $g=\gamma/(1-\kappa)$ and the magnification $\mu=1/\left((1-\kappa)^2-|\gamma|^2\right)$.

\subsection{CMB lensing}

\begin{figure}
\begin{center}
\psfig{figure=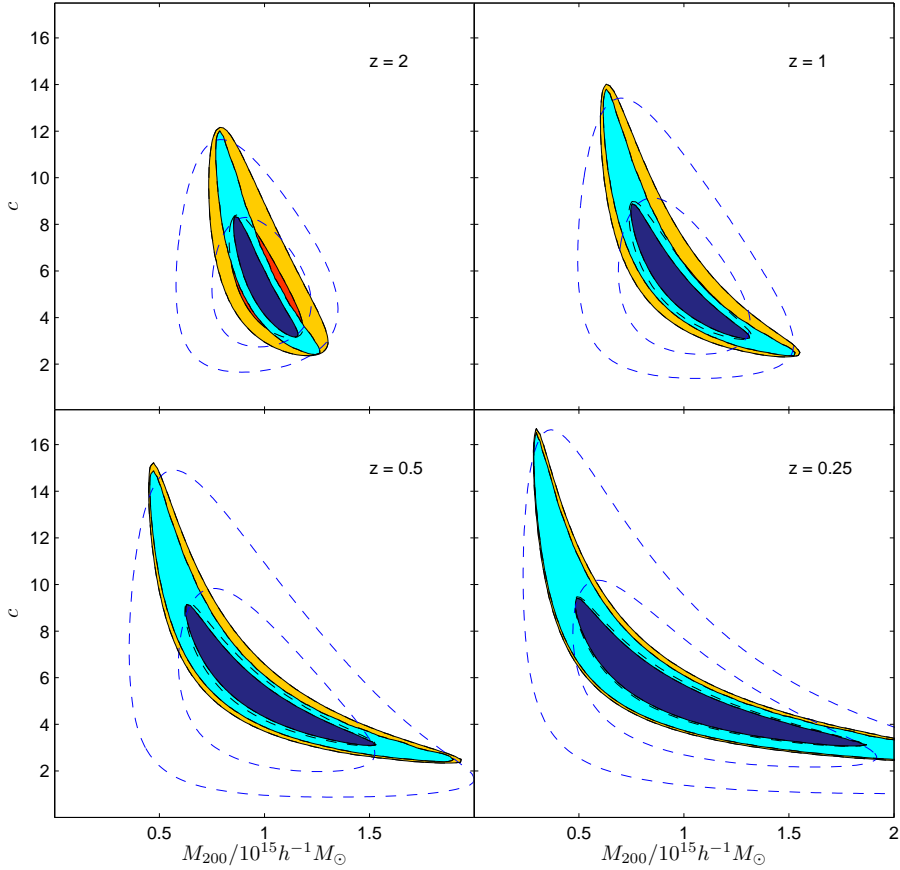,width=12cm}
\caption{Mean log likelihood constraints for a $M_{200} = 10^{15} h^{-1} M_\odot$ cluster with concentration parameter $c=5$ using CMB temperature lensing only. Dark solid is CMB lensing for $1\muK$ noise per $0.5\,\arcmin$ pixel assuming only lensing signal, paler solid is when a moving lens contribution is added and constraints are marginalized assuming a Gaussian velocity distribution with $\vrms = 300 \km s^{-1}$. Dashed lines show the constraints marginalized over an unknown rotating kinetic SZ contribution.
Contours show where the exponential of the mean log likelihood drops to 0.32 and 0.05 of the maximum.
\label{MoveCompare}}
\end{center}
\end{figure}

\begin{figure}
\begin{center}
\psfig{figure=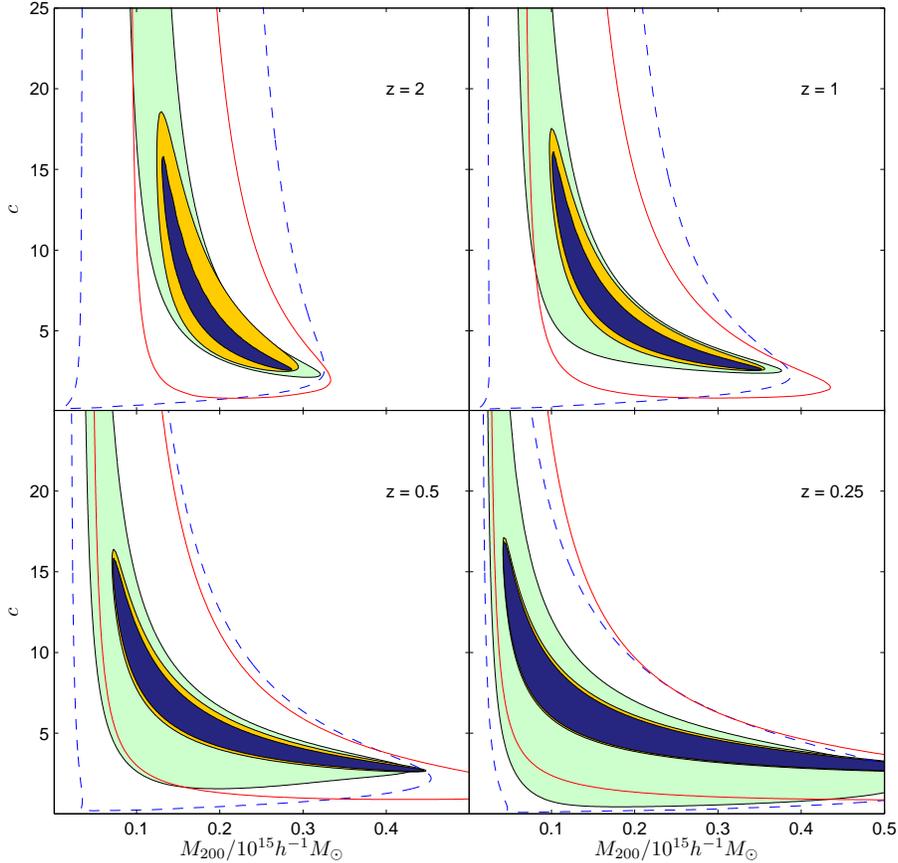,width=12cm}
\caption{Mean log likelihood constraints for a $M_{200} = 2\times 10^{14} h^{-1} M_\odot$ cluster with concentration parameter $c=5$ using CMB lensing only with noise $0.1\muK$ per $0.5\,\arcmin$ pixel. Filled contours are for temperature lensing, from the inside out (dark to light) they are: 1. CMB lensing only; 2. Marginalized over moving lens signal; 3. Marginalized over moving lens signal and including small scale power from inhomogeneous reionization. The dashed line shows the result when the kinetic SZ signal from cluster rotation is unknown and also marginalized. The solid unfilled contour shows the result from a clean polarization observation with noise $\sqrt{2}\times 0.1\muK$ on each Stokes' parameter. Contours enclose the area where the exponential of the mean log likelihood is greater than $0.05$ of the peak value.
\label{MoveCompareLowM}}
\end{center}
\end{figure}

The parameters we shall attempt to constrain are the concentration $c$ and the mass parameterized by $M_{200}$. For simplicity we shall assume the position of the centre is known (for example from observations of the thermal SZ), so there are only two parameters.

Since our model is spherically symmetric the kinetic SZ from line of sight motion is also symmetric about the cluster centre, and therefore orthogonal to the lensing signal. It can therefore be neglected. We show the effect of marginalizing over the various asymmetric signals on massive cluster constraints in Fig.~\ref{MoveCompare}. As a rough model of the moving lens signal we take a constant $\vrms=300\km s^{-1}$ (see e.g. Ref.~\cite{Colberg:1998ds}). Although only a fairly small effect at low redshift where the constraints are weak anyway, at high redshift the moving lens signal significantly increases the uncertainty in the cluster mass. Kinetic SZ from cluster rotation is a big problem if this cannot be modelled, and reflects the fact that temperature mass measurements are probably kinetic SZ limited~\cite{Holder:2004rp,Vale:2004rh}. At the $0.5 \muKarcmin$ noise level the polarization does not give a useful constraint due to the polarization signal being about ten times smaller than the temperature.

At high redshift there are expected to be almost no very massive clusters, so in Fig.~\ref{MoveCompareLowM} we show the effect for a more realistic cluster with $M_{200} = 2\times 10^{14} h^{-1} \Msun$ with ten times lower noise, $0.05\muKarcmin$ ($\sqrt{2}\times 0.05\muKarcmin$ on the Stokes' parameters). In the absence of confusing signals the temperature CMB gives excellent constraints at high redshift. However these are massively degraded by marginalization over other non-linear effects. The effect of more small scale power from inhomogeneous reionization significantly degrades the constraints at low redshift. We also show the constraint expected from a clean polarization observation, which at this noise level is competitive with what can realistically be achieved from the temperature.

\subsection{Galaxy lensing}

\begin{figure}
\begin{center}
\psfig{figure=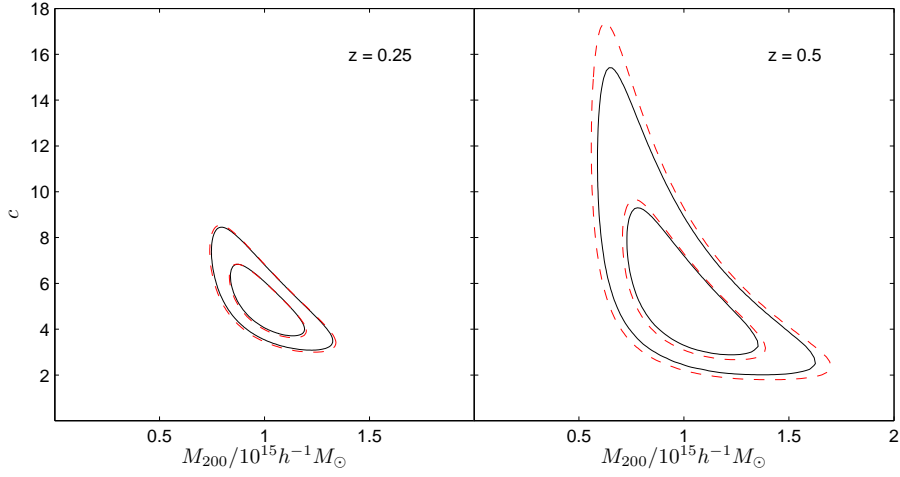,width=12cm}
\caption{Mean log likelihood constraints for a $M_{200} = 10^{15} h^{-1} \Msun$ cluster with concentration parameter $c=5$ from galaxy lensing with 30 galaxies arcmin$^{-2}$ and an intrinsic ellipticity dispersion $\sigma_u=0.3$. Solid contours show the constraint for $\la z \ra = 1$ when galaxy redshifts are known, dashed is the equivalent result when the redshift distribution is known. Contours show where the exponential of the mean log likelihood drops to 0.32 and 0.05 of the maximum.
\label{margecompare}}
\end{center}
\end{figure}

\begin{figure}
\begin{center}
\psfig{figure=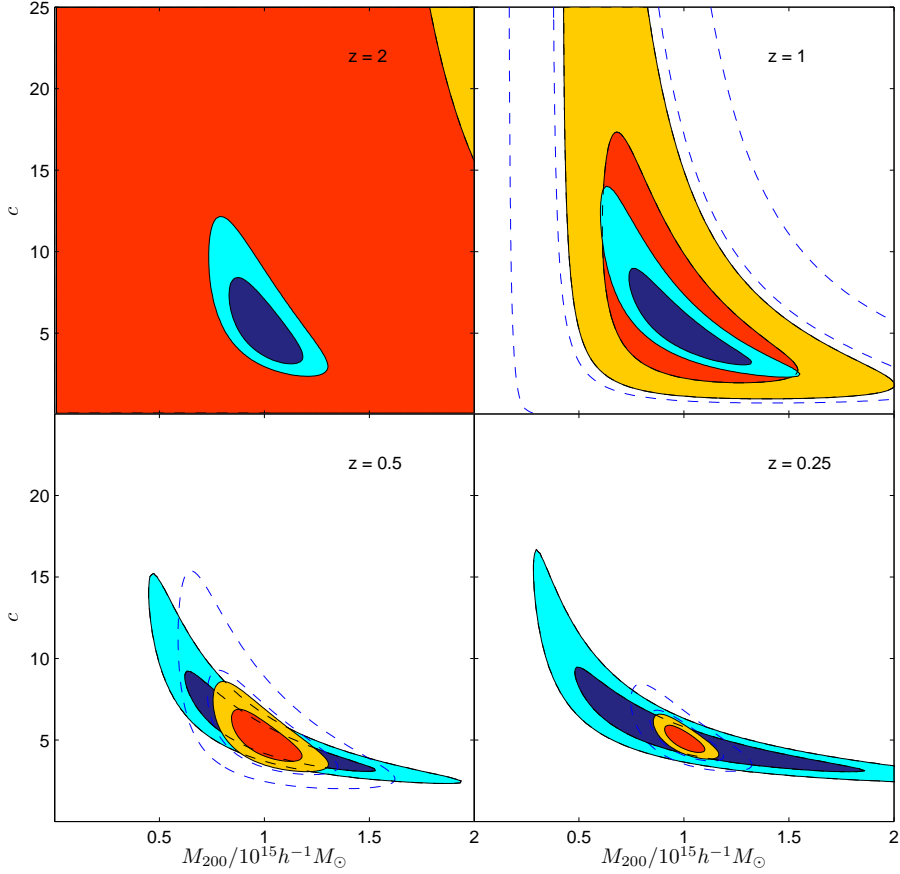,width=12cm}
\caption{Mean log likelihood constraints for a   $M_{200} = 10^{15} h^{-1} \Msun$ cluster with concentration parameter $c=5$. Dark solid is CMB lensing for $1\muK$ noise per $0.5\arcmin$ pixel marginalized over the moving lens signal, red solid is for space-based galaxy lensing with 100 galaxies arcmin$^{-2}$, dashed lines are for current ground-based galaxy lensing with 30 galaxies arcmin$^{-2}$. All galaxies have known redshifts, $\sigma_u=0.3$, and their distribution has $\la z \ra=1$. Contours show where the exponential of the mean log likelihood drops to 0.32 and 0.05 of the maximum.
\label{M1CMBCompare}}
\end{center}
\end{figure}

\begin{figure}
\begin{center}
\psfig{figure=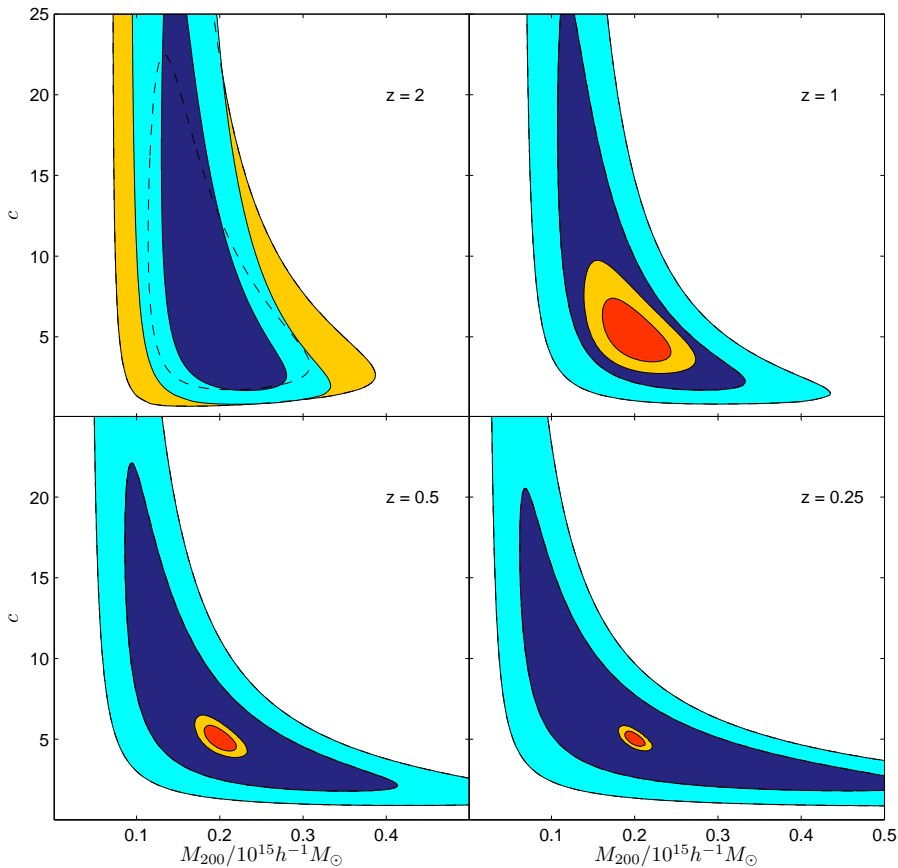,width=12cm}
\caption{Futuristic mean log likelihood constraints for a $M_{200} = 2\times 10^{14} h^{-1}\Msun$ cluster with concentration parameter $c=5$. Dark solid is for clean CMB polarization lensing for $\sqrt{2}\times 0.1\muK$ noise per $0.5\arcmin$ pixel on the Stokes' parameters, paler solid is for galaxy lensing with perfect redshift information and a total of 500 galaxies arcmin$^{-2}$ with a
Gaussian ellipticity distribution. All galaxies have $\sigma_u=0.2$ and the distribution has $\la z \ra=1.5$.
Contours show where the exponential of the mean log likelihood drops to 0.32 and 0.05 of the maximum.
\label{galcompare_m02_fut}}
\end{center}
\end{figure}

We assume that the sources are randomly distributed on the sky, and have a redshift probability distribution  $P(z){\rm d}z$ of the form suggested by Ref.~\cite{Brainerd:1995da}
\begin{equation}
P(z){\rm d}z=\frac{\eta\,z^2\exp\left[-\left(z/z_{0}\right)^\eta\right]}{\Gamma\left(\frac{3}{\eta}\right)z_{0}^3}\,{\rm d}z\,.
\end{equation}
Choosing $P(z){\rm d}z=(27/2)z^2\exp\left[-3z\right]{\rm d}z$ gives $\left<z\right>=1$, which is typical of current
deep surveys. For comparison, we also consider a survey with $\left<z\right>=1.5$, the form of which is $P(z){\rm d}z=9.57z^2\exp\left[-2.91z^{0.78}\right]{\rm d}z$. The distributions are skewed and peak at $z=2/3$ and $z\approx 0.85$ respectively.

For clusters with $z\lesssim 0.3$, and for typical (current) weak lensing observations with $\left<z\right>=1$, the cluster's lensing properties are quite insensitive to the actual redshift distribution of galaxies. The faint galaxy population can be safely approximated as lying on a sheet at $\left<z\right>$, or more precisely at the redshift corresponding to $\left<1/\Sigma_{\rm crit}(z)\right>$, with $\Sigma_{\rm crit}(z)$ being the critical surface mass density. We do not however make this approximation here. The availability of photometric redshifts for galaxies is becoming increasingly likely with the advent of wide-field infrared imagers to complement observations from optical instruments, and parameter estimates from such data sets would be comparable to those with spectroscopic redshifts~\cite{King:2000wc}. When there is no direct information about redshift, the galaxy's magnitude can be used to determine whether it is likely to be behind the cluster, so  $P(z'|z) \sim P(z')$ for $z'$ larger than about the cluster redshift, and small or zero otherwise.

For the $M_{200} = 2\times 10^{14} h^{-1} M_\odot$ cluster, the minimum radius of the aperture from which data is assumed to be available is 0.145\,Mpc, and for the $M_{200} = 10^{15} h^{-1} M_\odot$ cluster it is 0.35\,Mpc. For the outer radius, we integrate until the shear signal is negligible, and convergence in constraints is obtained. For the most massive cluster at $z=0.25$, this is equivalent to an angular scale of $\sim 15\,\arcmin$ (or to that of a wide-field imager such as that on the ESO-MPG 2.2m telescope).

Fig.~\ref{margecompare} shows expected constraints on massive cluster parameters from galaxy lensing with present day parameters. The constraints on clusters with $z\agt 1$ are very weak with current data, so we show only the useful constraints for lower redshift clusters. The figure shows the degrading effect of not knowing the source galaxy redshifts which have been marginalized over using Eq.~\ref{marge_eq}: for the $z=0.5$ cluster this is a noticeable effect, however for $z=0.25$ most of the galaxies lie well behind the cluster so knowing the individual redshifts gains very little. At higher redshifts the effect would be much more important, however future surveys able to give a useful constraint for higher redshift clusters would almost always have redshift information anyway.

In Fig.~\ref{M1CMBCompare} we compare current galaxy lensing constraints to those from possible future CMB temperature lensing observations. Even current galaxy lensing data can do better than futuristic CMB temperature lensing for low redshift clusters. However for $z\agt 0.5$ CMB lensing would be able to improve on current constraints if asymmetric kinetic SZ contamination and background secondaries did not destroy the signal. The fact that the CMB lensing constraints are weaker at low redshift when the deflection angle is larger may seem odd. However the point is that the deflection itself is unobservable, only the dipole-like variation can be cleanly distinguished from variations in the unlensed CMB. The constraints are therefore better for higher redshift clusters where the cluster has a smaller angular size, and hence the dipole-like pattern is more easily distinguished from larger scale variations in the unlensed CMB.

If Fig.~\ref{galcompare_m02_fut} we compare futuristic constraints from clean CMB polarization lensing with constraints from futuristic galaxy lensing with $\la z\ra=1.5$ (CMB polarization and temperature are compared in Fig.~\ref{MoveCompareLowM}). For clusters with redshifts well below the galaxy distribution peak, galaxy lensing should be much more powerful than CMB lensing.
At redshifts higher than the peak of the galaxy distribution function, the galaxy lensing results however become much worse due to the lack of sources, and CMB polarization lensing may be a better way to measure the mass of clusters at redshift $z \agt 1$. At these redshifts the CMB lensing result is also less degenerate with the concentration parameter than at low redshift.

Another source of noise in the determination of cluster mass profiles from galaxy lensing that we have neglected is the large scale structure along the line of sight. This has been estimated to cause as much as a factor of 2 increase in uncertainties in
estimates of $c$ and $M_{200}$ ~\cite{Hoekstra:2002cq}. Structure correlated with a cluster leads to a bias, with the masses of clusters in excess of $10^{14} h^{-1} M_\odot$ being overestimated by about 20\%~\cite{White:2001gs}. Techniques are being developed to alleviate these issues (e.g. see Ref.~\cite{Dodelson:2003tr}). For a particular cluster, if photometric or spectroscopic redshift information is available for the data field, then to some extent the importance of the large scale structure can be assessed by identifying any structures (e.g. galaxy groups) along the line of sight, or correlated large scale structure such as filaments.
\subsection{CMB lensing with non-symmetric profiles}

\begin{figure}
\begin{center}
\psfig{figure=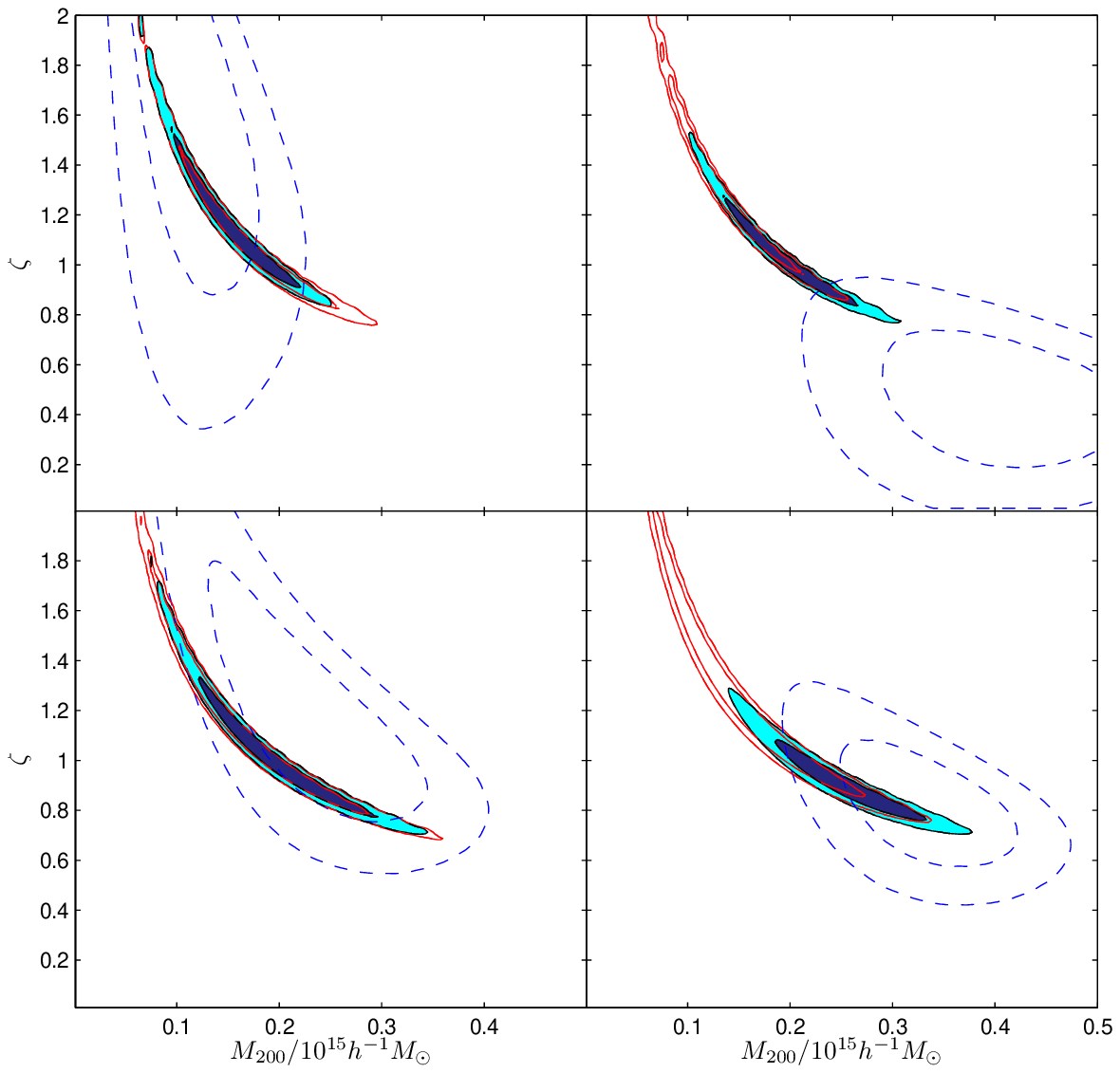,width=8.5cm}
\psfig{figure=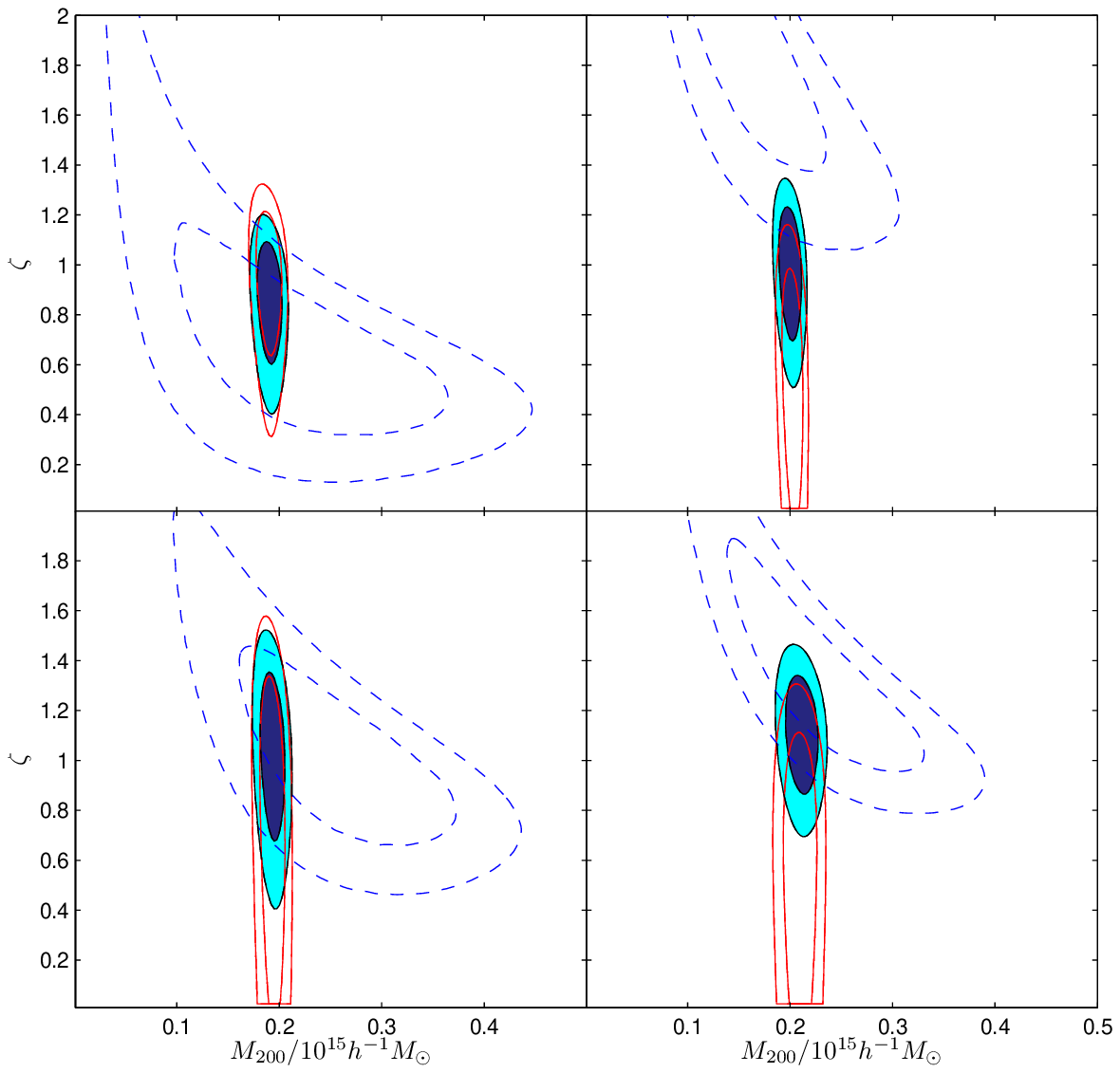,width=8.5cm}
\caption{Constraints on the stretch parameter $\zeta$ from four random realizations, for direction parallel (left) and orthogonal (right) to the direction of the background CMB temperature gradient. Solid lines are using CMB temperature only, dashed polarization only, filled contours show the combined result (accounting for the correlation). The fiducial cluster was at redshift one and has a spherically symmetric NFW profile with mass of $M_{200} = 2\times 10^{14} h^{-1} \Msun$. The noise is $0.1\muK$ per $0.5\,\arcmin$ pixel ($\sqrt{2}\times 0.1\muK$ for the polarization) the concentration is fixed to $c=5$.
\label{stretch}}
\end{center}
\end{figure}

So far we have presented results for mean log likelihoods that give an idea of average results per cluster when you observe many. However one must remember that CMB temperature lensing constraint is highly variable due to the variability of the gradient behind any given cluster. If a cluster happens to be in the middle of a broad hot spot there will be essentially no constraint.

The assumption of spherical cluster symmetry is also qualitatively important for the CMB temperature lensing results. If a cluster is observed in front of a pure gradient field, then deflection in the direction orthogonal to the gradient will be unconstrained. If we assume spherical symmetry this does not really matter as the profile can be constrained from the gradient-direction deflections. However if the cluster is non-symmetric, there is a fundamental degeneracy: CMB lensing of a pure gradient field cannot constrain orthogonal deflections at all. Of course the CMB is not a pure gradient, and in this case small scale CMB power actually helps to break this degeneracy to some extent. Furthermore CMB polarization provides two additional gradient directions, so the probability of all three gradients being aligned or very small is low. Polarization observations should be able to significantly help general cluster profile reconstruction. This is important because in reality the majority of clusters are expected to be prolate~\cite{Shaw:2005dy} rather than spherical, as well as having interesting substructure.

General cluster profile reconstruction from CMB lensing is beyond the scope of this paper (see Ref.~\cite{Maturi:2004zj} for some work with CMB temperature). Here we consider a simple toy model to illustrate the issues: we try to reconstruct a stretched NFW deflection field
\begin{equation}
\valpha = \zeta \hat{\vn}\cdot \valpha_{\text{NFW}}\, \hat{\vn}  + \hat{\vn}_\perp \cdot \valpha_{\text{NFW}} \,\hat{\vn}_\perp
\end{equation}
where $\zeta$ is a stretch parameter. We choose $\hat{\vn}$ to be in the direction of the CMB temperature gradient or the direction orthogonal to it, where $\hat{\vn}_\perp$ is the corresponding orthogonal direction. For simplicity we fix the concentration parameter to $c=5$, and only consider the idealized case where the temperature signal is purely CMB lensing. Note that this toy model is not a realistic lensing deflection angle field as in general it is not curl free.

In Fig.~\ref{stretch} we show the constraints on the stretch in the two orthogonal directions for four different realizations of the underlying fields. The CMB temperature stretch constraint in the direction orthogonal to the CMB gradient is generally very poor. The CMB polarization constraints are slightly correlated, but generally in significantly different directions, meaning that joint constraints are significantly better than those from the CMB temperature alone.

The presence of smaller scale CMB power from e.g. inhomogeneous reionization can actually improve the constraints from the temperature in the orthogonal direction due to the extra information available in the smaller scale power. However, as emphasised above, the CMB temperature signal is likely to be complicated due to SZ etc, so in practice the CMB temperature result is likely to be very much worse than that shown here. CMB polarization may be a much better probe.

\section{Conclusions}

Weak lensing is a valuable and promising method for studying cluster masses. Measurements of lensing shear using observations of lensed galaxies provides tight constraints for low redshift clusters. For clusters at higher redshift than the peak of the galaxy-redshift distribution the galaxy lensing constraints become much poorer, and CMB lensing can do better. However, even with simple spherically symmetric models the temperature lensing signal can be degraded by various other second order effects. For futuristic arcminute-resolution observations at low noise levels the CMB polarization lensing signal may be much cleaner and a more robust way to measure cluster properties. Measurements of lensing by high redshift clusters is therefore something that future CMB polarization missions may wish to aim to achieve.

The results from galaxy lensing are limited by the intrinsic ellipticity dispersion of the galaxies, and the fact that there are only a finite number of sources behind the cluster. To do better one could try to find sources which have a higher number density. Possibilities include high redshift sources observed with 21cm, sources from the time of inhomogeneous reionization, and secondary doppler CMB signals from velocities after reionization~\cite{Zaldarriaga:1998te,Pen:2003yv,Zahn:2005ap}. In addition strong lensing can be used to help constrain the central region of the cluster profile.
The ultimate limit from CMB polarization will depend on how efficiently spectral information can be used to clean out confusing signals, and the extent to which cluster substructure complicates the signal from quadrupole scattering. Future work could investigate this using numerical simulations.

With the appropriate increase in resolution and sensitivity, methods for cluster CMB lensing could be extended to constrain galaxy profiles (as discussed for the CMB temperature in Ref.~\cite{Dodelson:2003gv}).

\section{Acknowledgments}

We thank Anthony Challinor, Jochen Weller for helpful suggestions, also Peter Schneider and Richard Battye. We thank CITA for use of their Beowulf computer. AL acknowledges support via a PPARC Advanced Fellowship. LJK thanks the Royal Society for support via a University Research Fellowship.

\appendix

\section{Gradient approximation}

Since gradients are just linear combinations of the (assumed) Gaussian underlying fields, the gradient fields are also Gaussian and hence fully described by their covariance. We can either work out the full distribution of the gradients directly, or just work out the covariances. Here we chose to do the latter, and calculate the covariances of the gradients at a point assuming statistical isotropy. The temperature gradient variance is given by
\begin{eqnarray}
2P_{\Theta} \equiv \la |\grad \Theta|^2\ra &=& \la \bigl|\sum_{lm} \Theta_{lm} \grad Y_{lm}\bigr|^2 \ra\nonumber \\
&=& -\sum_l C^\Theta_l \frac{1}{4\pi} \sum_m \int \ud\Omega Y_{lm} \grad^2 Y_{lm}^*  \nonumber\\
&=&\sum_l l(l+1) \frac{2l+1}{4\pi}  C_l^\Theta,
\end{eqnarray}
where we used statistical isotropy and orthogonality of the spherical harmonics.
A polarization tensor $P_{ab}$ can be defined so that in a fixed orthonormal basis
\begin{equation}
P_{ij} =\frac{1}{2} \begm Q & U \\ U & -Q \enm,
\end{equation}
and may be expanded in terms of gradient and curl tensor spherical harmonics~\cite{Kamionkowski:1996ks} $Y_{ab}^G{}_{(lm)}$ and $Y_{ab}^C{}_{(lm)}$. The harmonic components describe the $E$- and $B$-modes of the polarization respectively.
The correlation of the polarization divergence and temperature is given by
\begin{eqnarray}
P_X \equiv \frac{1}{2}\la \grad^a P_{ab} \grad^b T\ra &=& \frac{1}{2\sqrt{2}}\sum_{lm} C_l^X \frac{1}{4\pi} \int \grad^a Y_{ab}^G{}_{(lm)} \grad^b Y_{lm}^* \nonumber\\
&=& - \frac{1}{4} \sum_l \sqrt{\frac{(l+2)!}{(l-2)!}} \frac{2l+1}{4\pi} C_l^X,
\end{eqnarray}
where $C_l^X$ is the temperature-polarization cross-correlation power spectrum.
The other terms are zero:
\begin{equation}
\la \grad^a P_{ab}\, \epsilon^b{}_c\grad^c \Theta \ra = 0.
\end{equation}
The variance of the polarization divergence is
\begin{equation}
P_P\equiv \la \grad^a P_{ab} \grad^c P_c{}^b \ra = \frac{1}{4}\sum_l (l+2)(l-1) \frac{2l+1}{4\pi}  (C_l^E+ C_l^B).
\end{equation}
Similar results can be derived for the variance of the irreducible polarization gradient 3-tensor.
In terms of a fixed flat-sky basis we have
\begin{equation}
\la \grad_x U \grad_y \Theta\ra = \la \grad_y U \grad_x \Theta\ra = \la \grad_x Q \grad_x \Theta\ra = -\la \grad_y Q \grad_y \Theta\ra = P_X,
\end{equation}
\begin{equation}
\la |\grad Q|^2\ra = \la |\grad U|^2\ra = 2 P_P,
\end{equation}
with other terms being zero, as can readily be verified by using an explicit flat-sky harmonic expansion. The stochastic quantities are $(\grad_x \Theta, \grad_y \Theta, \grad_x Q, \grad_y Q,\grad_x U, \grad_y U)$, which therefore have covariance
\begin{equation}
\begm
P_T & 0 & P_X & 0 & 0 & P_X \\
0 & P_T & 0 & P_X & -P_X & 0 \\
P_X & 0 & P_P & 0 & 0 & 0 \\
0 & P_X & 0 & P_P & 0 & 0 \\
0 & -P_X & 0 & 0 & P_P & 0 \\
P_X & 0 & 0 & 0 & 0& P_P
\enm.
\end{equation}
The correlation $P_X$ is negative, and the (anti-)correlation $P_X/(P_\Theta P_P)^{1/2}$ is about $10\%$. On the small scales we are considering, photons are flowing into hot spots, which for a plane wave means that the correlated part of $Q$ (defined with respect to the wavevector---parallel to $\grad \Theta$---basis direction) is negative on the the hot crest~\cite{Coulson:1994qw}. $Q$ therefore increases in the direction of the cold crest, and hence that the gradients in $Q$ and $\Theta$ are anti-correlated.
Note for our conventions we have to flip sign of $C_l^X$ from CAMB/CMBFAST which use the conventions of Ref.~\cite{Zaldarriaga:1996xe}.

In a fixed flat sky basis, using the gradient approximation,  some deflection field $\valpha(\vr)$ gives the lensed polarization field
\begin{equation}
\tilde{P}_{ab}(\vr) = P_{ab}(\vr) +  \valpha(\vr) \cdot \vgrad P_{ab}.
\end{equation}
A constant gradient is neither $E$ or $B$-mode,  because making the $E$/$B$ decomposition locally requires taking two gradients. This is reflected in the fact that the covariance is a function only of the sum of the power spectra $C_l^E+C_l^B$.  On the flat sky the scalar harmonics are $e^{i\vl\cdot\vx}$ and tensor harmonics are (we follow the conventions of Ref.~\cite{Lewis:2006fu})
\begin{eqnarray}
Q^G_{ab}(\vl) &=& - \sqrt{2}\,\, \vlhat_{\la a} \vlhat_{b\ra}  e^{i\vl\cdot \vx} \\
Q^C_{ab}(\vl) &=& - \sqrt{2}\,\, \epsilon^c{}_{(a} \vlhat_{b)}  \vlhat_c  e^{i\vl\cdot \vx},
\end{eqnarray}
where angle brackets denote the symmetric trace free part of the enclosed indices and $\vlhat=\vl/|\vl|$. Assuming a suitably well behaved deflection field the lensed harmonic components are then
\begin{eqnarray}
\tE(\vl) &=& -2\valpha(\vl) \cdot \vgrad P_{ab}\,\vlhat^{\la a} \vlhat^{b\ra}  \nonumber\\
\tB(\vl) &=& -2\valpha(\vl) \cdot \vgrad P_{ab}\, \epsilon_c{}^{(a} \vlhat^{b)}\vlhat^c.
\label{cluster_EB}
\end{eqnarray}
Averaging over (assumed Gaussian) background polarization gradients $\vgrad P_{ab}$ we get
\begin{eqnarray}
\la |\tE(\vl)|^2\ra &=& P_P |\valpha(\vl)|^2 \nonumber\\
\la |\tB(\vl)|^2\ra&=& P_P|\valpha(\vl)|^2.
\end{eqnarray}
 An arbitrary deflection field therefore gives identical power for each $E$ and $B$ mode on average when lensing a pure gradient field. If the deflection field has circular symmetry (as from a spherical cluster), the angular average of the $E$ and $B$ mode power are equal for any fixed polarization gradient:
\begin{eqnarray}
\frac{1}{2\pi}\int \ud\phi_\vl |\tE(\vl)|^2 &=& \frac{\alpha^2(l)}{4}\left(|\vgrad Q|^2 + |\vgrad U|^2\right)\nonumber\\
\frac{1}{2\pi}\int \ud\phi_\vl |\tB(\vl)|^2 &=& \frac{\alpha^2(l)}{4}\left(|\vgrad Q|^2 + |\vgrad U|^2\right).
\end{eqnarray}

Cluster lensing therefore generates equal amplitude $E$ and $B$ in the gradient approximation.
However there is very little power in the unlensed $E$ or $B$ polarization on cluster scales,
so the lensed $E$ contains almost as much information as the lensed $B$.
We use the full likelihood function so there is in fact no need to use $E$ and $B$. However for nearby large clusters the $B$ mode signal may allow the cluster signal to be distinguished from unlensed CMB `noise', so unlike in the temperature case cluster lensing may not be CMB noise limited for large cluster sizes.

Since the temperature gradient defines a direction, we could chose to define the Q and U Stokes' parameters with respect to this variable basis, e.g. where we align the x-axis with $\grad \Theta$. In this basis
\begin{equation}
\la |\grad_x \Theta|^2\ra' = \la |\grad \Theta|^2\ra = 2 \la |\grad_x \Theta|^2 \ra,
\end{equation}
simply twice the variance of the x-component in a fixed basis, and similar results hold for the non-zero terms. By choosing this basis we are however making the whole distribution non-Gaussian, for example the marginalized distribution of $\delta\equiv (\grad_x \Theta)' = |\grad \Theta|$ is $P(\delta) \propto \delta\exp(-\delta^2/2P_T)$.


\providecommand{\apj}{Astrophys. J. }\providecommand{\apjl}{Astrophys. J.
  }\providecommand{\mnras}{MNRAS}

\end{document}